\documentclass[12pt]{article}
\usepackage{amssymb,graphicx,amsmath,array,verbatim,cite}
\makeatletter

\@addtoreset{equation}{section}
\makeatother
\renewcommand{\thefootnote}{\arabic{footnote}}

\newcommand{\p}{\partial}
\newcommand{\beq}{\begin{eqnarray}}
\newcommand{\eeq}{\end{eqnarray}}
\newcommand{\vs}[1]{\vspace{#1 mm}}
\newcommand{\hs}[1]{\hspace{#1 mm}}
\newcommand{\bpm}{\begin{pmatrix}}
\newcommand{\epm}{\end{pmatrix}}
\newcommand{\Z}{\mathbb{Z}}
\newcommand{\R}{\mathbb{R}}
\newcommand{\C}{\mathbb{C}}

\newcommand{\tr}{{\rm Tr}}
\newcommand{\D}{\mathcal D}
\newcommand{\ba}{\left( \begin{array}}
\newcommand{\ea}{\end{array} \right)}

\usepackage[usenames]{color}

\begin{document}
\renewcommand{\thefootnote}{\fnsymbol{footnote}}

\title{{\bf Instantons in Lifshitz Field Theories}}
\author{Toshiaki Fujimori and Muneto Nitta 
\\{\it\small 
Department of Physics, and Research and Education Center for Natural Sciences,} 
\\{\it\small 
Keio University, Hiyoshi 4-1-1, Yokohama, Kanagawa 223-8521, Japan}}

\maketitle

\vs{15}
\begin{abstract}
BPS instantons are discussed in Lifshitz-type anisotropic field theories. 
We consider generalizations of the sigma model/Yang-Mills instantons 
in renormalizable higher dimensional models with the classical Lifshitz scaling invariance. 
In each model, BPS instanton equation takes the form of the gradient flow equations 
for ``the superpotential" defining ``the detailed balance condition". 
The anisotropic Weyl rescaling and the coset space dimensional reduction 
are used to map rotationally symmetric instantons to vortices 
in two-dimensional anisotropic systems on the hyperbolic plane.
As examples, we study anisotropic BPS baby Skyrmion 
$1+1$ dimensions and BPS Skyrmion in $2+1$ dimensions, 
for which we take K\"ahler 1-form and 
the Wess-Zumiono-Witten term as the superpotentials, respectively, 
and an anisotropic generalized Yang-Mills instanton in $4+1$ dimensions,
for which we take the Chern-Simons term as the superpotential.
\end{abstract}

\newpage
\tableofcontents
\newpage 

\section{Introduction}
Instantons are important objects 
which are relevant to non-perturbative phenomena in various physical system.   
They are characterized as topologically non-trivial field configurations 
which are stationary points of the Euclidean action of the system, 
and hence they give non-perturbative contributions to path integrals.  
Well-known examples of instantons include 
the sigma model instantons in two dimensional non-linear sigma models and 
the Yang-Mills instantons in four dimensional non-Abelian gauge theories. 
Both models are renormalizable asymptotically free quantum field theories
which have rich non-perturbative structures in the low-energy regime. 

Although the straightforward generalizations of those theories 
to higher dimensions are known to be perturbatively non-renormalizable, 
there are renormalizable higher dimensional generalizations of the sigma model and the gauge theory. 
Such theories, called the Lifshitz-type fields theories, 
are characterized by the following anisotropic scaling of the spacetime coordinates \cite{Lifshitz}
\beq
t \rightarrow \lambda^z t, \hs{10} x_i \rightarrow \lambda x_i. 
\label{eq:scaling}
\eeq
This scaling transformation is called the Lifshitz scaling with the dynamical critical exponent $z$. 
Because of the anisotropy between space and time, 
the Lorentz symmetry is explicitly broken, 
whereas the spatial rotational symmetry is preserved. 
The spatial higher derivative terms in the Lagrangian
improve the ultraviolet (UV) behaviors of the propagators 
so that the UV divergences of loop integrals become milder than 
those in the standard Lorentz symmetric theories.  
The absence of higher time derivative terms guarantees that 
the Lifshitz-type theories are free from the ghost problem associated with higher derivatives. 

Using the idea of the Lifshitz scaling, 
Horava constructed a gravity theory \cite{Horava:2008ih,Horava:2009uw}, 
called ``the Horava-Lifshitz gravity", which is expected to be renormalizable and unitary.
In $(d+1)$ dimensional spacetime, 
non-linear sigma models with $z=d$ and 
Yang-Mills theories with $z=d-2$ 
are classically invariant under the Lifshitz scaling transformation,  
and expected to be renormalizable according to the weighted power counting \cite{Anselmi:2007ri,Anselmi:2008bq,Anselmi:2008bs}
and a symmetry argument \cite{Fujimori:2015wda}. 
The various aspects of the Lifshitz-type theories have been discussed  
in the non-linear sigma models \cite{Anselmi:2008ry, Das:2009ba, Anagnostopoulos:2010gw, Gomes:2013jba} 
and in the Yang-Mills theories \cite{Horava:2008jf, Chen:2009ka, Kanazawa:2014fla} (see \cite{Alexandre:2011kr} for a review). 
In most of those cases, the theories are asymptotically free (or scale invariant)
and it is likely that their strongly coupled low-energy dynamics  
have rich non-perturbative structures induced by instantons. 

In recent years, BPS solitons has been extensively studied 
in a certain class of Lorentz symmetric non-linear sigma models with higher derivative terms 
\cite{MullerKirsten:1990qw,Gisiger:1996vb,Adam:2009px,Adam:2010fg,
Adam:2010jr,Speight:2010sy,Adam:2010ds,Bonenfant:2010ab,Adam:2012aq,
Adam:2012sa,Adam:2013hza,Adam:2011hj,Adam:2013awa,
Nitta:2014pwa,Nitta:2015uba,Bolognesi:2014ova,Adam:2014xfa}. 
Those models are invariant under the volume-preserving diffeomorphism, 
which allows one to determine various aspects of solitons. 
It is also possible to implement such an infinite dimensional symmetry 
in the Lifshitz-type non-linear sigma models 
as the spatial volume-preserving diffeomorphism.
BPS solitons in such Lifshitz-type models were discussed in \cite{Adam:2012dg}
(another kind of BPS solitons in Lifshitz field theories have also been discussed in \cite{Kobakhidze:2010cq}).

In this paper, we discuss instantons in classically scale invariant 
Lifshitz-type non-linear sigma models and gauge theories in the Wick-rotated spacetime. 
We focus on ``supersymmetric theories" which satisfy ``the detailed balance condition" 
characterized by a functional version of the superpotential.  
Such models admit BPS instantons described by the gradient flow equations for the superpotential
(similar geometric flow equations have been discussed 
as instantons in the Horava-Lifshitz gravity \cite{Bakas:2010fm, Bakas:2010by}).
A class of non-linear sigma models which is invariant under 
the Lifshitz scaling and the spatial volume-preserving diffeomorphism
can be constructed by adopting closed forms on the target space as the superpotential. 
As examples we study an anisotropic baby Skyrmion
in 1+1 dimensions and BPS Skyrmion in 2+1 dimensions, 
for which we take the K\"ahler 1-form and 
the Wess-Zumino-Witten term \cite{Witten:1983ar} as the superpotentials, respectively. 
For gauge theories, we use the Chern-Simons form to construct scale invariant models. 
In both cases, we consider rotationally invariant instantons 
and reduce the system to the two-dimensional half-plane endowed with the hyperbolic metric. 

The organization of this paper is as follows. 
In Sec.\,\ref{sec:sigma_model}, we discuss BPS instantons in the Lifshitz-type non-linear sigma model. 
After generalizing Derrick's scaling argument 
\cite{Derrick:1964ww} to models without the Lorentz symmetry in Sec.\,\ref{sec:Derrick}, 
we determine the Lifshitz-type sigma model action which admits BPS instantons in Sec.\,\ref{sec:action}. 
As examples, we consider anisotropic BPS (baby) Skyrmions in Secs.\,\ref{sec:baby} and \ref{sec:Skyrme}. 
In Sec.\,\ref{sec:higher}, generalizations to higher dimensional sigma models are discussed. 
Sec.\,\ref{eq:gauge} is devoted to the instantons in the Lifshitz-type gauge theories. 
In Sec.\,\ref{sec:Derrick_gauge}, we generalize the scaling argument to the Lifshitz-type gauge theories. 
In Sec.\,\ref{sec:red}, the reduction of the system 
by using the anisotropic Weyl rescaling  
and the coset space dimensional reduction are discussed. 
An explicit example of the generalized Yang-Mills instanton 
in $(5+1)$-dimensional $SU(4)$ gauge theory is discussed in Sec.\,\ref{sec:SU(4)}. 
Appendix \ref{appendix:SUSY} is devoted to the superfield formalism for 
the $(2+1)$-dimensional Lifsthiz-type supersymmetric non-linear sigma model.

%%%%%%%%%%%%%%%%%%%%%%%%%%%%%%%%%%%%%%%%%%%%%%
\section{Instantons in Lifshitz-type sigma models}
\label{sec:sigma_model}
In this section, we consider instantons in the Lifshitz-type non-linear sigma models. 
The degrees of freedom is a map from the $(d+1)$-dimensional spacetime to a target manifold $\mathcal M$
and the scalar fields $\phi^a$ are identified with the coordinates of $\mathcal M$. 
The generic form of the Euclidean action is 
\beq
S = \int dt d^d x \, \Big[ g_{ab} \p_t \phi^a \p_t \phi^b + V(\phi, \p_i \phi, \cdots) \Big], 
\label{eq:action_0}
\eeq
where $g_{ab}$ is the metric on $\mathcal M$ and 
$V(\phi, \p_i \phi, \cdots)$ is a function of the fields $\phi^a$ 
and their spatial derivatives $\p_i \phi^a, \p_i \p_j \phi^a$, etc. 
The model of this type is renormalizable if $V(\phi, \p_i \phi, \cdots)$ 
contains a sufficient number of spatial derivatives 
so that the propagator decreases sufficiently quickly in the UV regime. 
In the following, we discuss instantons in such models with higher spatial derivatives. 

%%%%%%%%%%%%%%%%%%%%%%%%%%%%%%%%%%%%%%%%%%%%%%
\subsection{Derrick's scaling argument for Lifshitz-type sigma models}
\label{sec:Derrick} 
First, we discuss the stability of instantons in the Lifshitz-type theories by using Derrick's scaling argument \cite{Derrick:1964ww}. 
Let us consider the following scaling transformation 
\beq
\phi^a(x_i) ~\rightarrow~ \phi^a(\lambda^{w_i} x_i), 
\label{eq:Gscaling}
\eeq
where $x_0 = t$ and $x_i$ are the coordinates in the Euclidean time and the spatial directions, respectively. 
If the action has non-trivial extrema, 
it should be stable under arbitrary variations of the fields. 
Thus, we can find a necessary condition for the existence of non-trivial extrema 
by looking at how the action behaves under the scaling Eq.\,\eqref{eq:Gscaling} 
with an arbitrary assignment of the scaling weights $w_i$.

Let $S_{(n_0, n_1, \cdots, n_d)}$ be the part of the action with $\#(\p_i) = n_i$.  
For example, the standard kinetic term with two time derivatives 
(the first term in Eq.\,\eqref{eq:action_0}) is denoted by $S_{(2,0,\cdots,0)}$. 
Under the transformation Eq.\,\eqref{eq:Gscaling}, $S_{(n_0, n_1, \cdots, n_d)}$ scales as
\beq
S_{(n_0,\cdots,n_d)} \rightarrow \lambda^{w_{(n_0, \cdots, n_d)}} S_{(n_0,\cdots,n_d)}, \hs{10} 
w_{(n_0, \cdots, n_d)} \equiv \sum_{i=0}^d (n_i-1) w_i. 
\eeq
Here we assume that each term $S_{(n_0,\cdots,n_d)}$ is finite and positive definite. 
Then the stability of the action under this scaling requires that 
there should be at least one pair of terms whose scaling weights have opposite signs. 
This is the necessary condition for the existence of non-trivial solutions of the equations of motion.

Let us first recall the Lorentz symmetric case in two dimensions.  
For a generic assignment of the scaling weights, 
the standard kinetic terms $S_{(2,0)}$ and $S_{(0,2)}$ have the weights with opposite signs
\beq
w_{(2,0)} = - w_{(0,2)} = w_0 - w_1.
\eeq
Thus, the Lorentz symmetric scalar field theories in two dimensions 
can have non-trivial solutions\footnote{
Note that Derrick's theorem does not guarantee the existence of non-trivial solutions}
which are stable under the transformations with $w_0 \not = w_1$. 
On the other hand, for $w_0=w_1$, both $S_{(2,0)}$ and $S_{(0,2)}$ are scale invariant, 
so that the non-trivial configurations are marginally stable under the scaling with $w_0 = w_1$. 
In other words, they have scale moduli which parameterize the set of marginally stable solutions. 
Such configurations with scale moduli are known as the sigma model instantons (a.k.a the sigma model lumps). 
Their size is fixed when there are additional terms with opposite scaling weights in the action. 
The baby Skyrmions are such solitons with fixed sizes in the models 
with potential terms $S_{(0,0)}$ and higher derivative terms $S_{(m,n)}$ 
\cite{Piette:1994ug,Piette:1994mh}.

We can generalize the above discussion to arbitrary dimensions. 
In general, $S_{(2,0,\cdots,0)}$ and $S_{(0,2,\cdots,2)}$ have the scaling weights with opposite signs
\beq
w_{(2,0,\cdots,0)} = - w_{(0,2,\cdots,2)} = w_0 - \sum_{i=1}^d w_i. 
\eeq
Thus, they are stable under the generic scaling transformations and 
invariant under any scaling transformaions with $w_0 = \sum_{i=1}^d w_i$. 
Therefore, the action of the form
\beq
S = S_{(2,0,\cdots,0)} + S_{(0,2,\cdots,2)},
\label{eq:stableaction}
\eeq 
can have non-trivial configurations with multiple scale moduli. 
Such scale moduli are fixed if there are additional potential and higher derivative terms. 
In the following, we mainly consider instantons in scale invariant theories 
with the action of the form of Eq.\,\eqref{eq:stableaction}.

%%%%%%%%%%%%%%%%%%%%%%%%%%%%%%%%%%%%%%%%%%%%%%
\subsection{``Supersymmetric" models and BPS equations}
\label{sec:action}
In this section, we discuss the form of the Lifshitz-type action which admits instanton solutions. 
We focus on the BPS case in which the action is bounded below by the topological charge of instantons. 
Such an action, which satisfies the so-called ``detailed balance condition", 
is characterized by a functional defined on each time slice
\beq
W = \int d^d x \, \mathcal W(\phi,\p_i \phi,\cdots), 
\eeq
where $\mathcal{W}$ is a function of $\phi^a$ and their spatial derivatives. 
In terms of the functional $W$, 
the action of the non-linear sigma model which admits BPS instantons is given by
\beq
S = \int dt d^d x \left[ \frac{1}{2} g_{ab} \p_t \phi^a \p_t \phi^b 
+ \frac{1}{2} g^{ab} \frac{\delta W}{\delta \phi^a} \frac{\delta W}{\delta \phi^b} \right] ,
\label{eq:action1}
\eeq
where the variation of $W$ takes the form
\beq
\frac{\delta W}{\delta \phi^a} ~=~ \frac{\p \mathcal W}{\p \phi^a} - \p_i \frac{\p \mathcal W}{\p \p_i \phi^a} + \{\mbox{higher derivative terms}\}. 
\eeq
For simplicity, we assume that $\mathcal W$ does not contain the higher derivative of $\phi^a$ in the following. 
In $(d+1)$-dimensional spacetime with $d=0,1,2$, 
the action Eq.\,\eqref{eq:action1} can be embeded into 
a supersymmetric model (see Appendix \ref{appendix:SUSY}) 
which has a complex supercharge $Q$ satisfying the algebra 
\beq
\{Q,\bar{Q}\} = 2 i \p_t, \hs{10} Q^2 = \bar{Q}^2=0.
\eeq
We call the functional $W$ a ``superpotential" since it is a generalization of 
the superpotential in the standard supersymmetric theories.
The supersymmetry in the Lifshitz-type field theory in higher dimensions will be discussed elsewhere. 

The action Eq.\,\eqref{eq:action1} can be rewritten into the Bogomol'nyi form
\beq
S ~=~ T + \int dt d^d x \, \frac{1}{2} g_{ab} \left( \p_t \phi^a+ g^{ac} \frac{\delta W}{\delta \phi^c} \right) 
\left( \p_t \phi^b + g^{bd} \frac{\delta W}{\delta \phi^d} \right) ~ \geq ~ T, 
\label{eq:bpsbound}
\eeq
where the topological charge $T$ takes the form
\beq
T =  \int dt d^d x \left[ - \p_t \mathcal W + \p_i \left( \p_t \phi^a \frac{\p \mathcal W}{\p \p_i \phi^a} \right) \right]. 
\eeq
Since the topological charge $T$ is a total derivative term, 
it has a fixed value for a given boundary condition. 
Therefore, the configurations with the least action 
in a fixed topological sector are the solutions of the following gradient flow equation
\beq
\p_t \phi^a = - g^{ab} \frac{\delta W}{\delta \phi^b} .
\eeq
This is the BPS equation which describes instantons in the Lifshitz-type sigma model.

Among various choices of the superpotential $W$, 
we focus on a specific class of $W$ which has several special symmetries. 
Let $\omega$ be a closed $(d+1)$-form on the target manifold $\mathcal M$
\beq
\omega = \omega_{a_1 \cdots a_{d+1}} d \phi^{a_1} \wedge \cdots \wedge d \phi^{a_{d+1}}, \hs{10} d \omega = 0.
\eeq
Then, we can find a functional $W$ such that\footnote{
Since $\omega$ is a closed form, the superpotential  
\beq
W ~=~ \int d^d x \int_{-\infty}^t dt' \ \omega_{a b_1 \cdots b_d} \p_{t'} \phi^a \p_{x_1} \phi^{b_1} \cdots \p_{x_d} \phi^{b_d}, \notag
\eeq
depends only on the field on the boundary. 
If we assume that the fields $\phi^a$ approaches to a single point on the target space $\mathcal M$,
the superpotential $W$ can be viewed as a functional defined on the time slice at $t$ 
whose variation gives Eq.\,\eqref{eq:variation}. 
}
\beq
\frac{\delta W}{\delta \phi^a} = \omega_{a \, b_1 \cdots b_d} \p_{x_1} \phi^{b_1} \cdots \p_{x_d} \phi^{b_d}.
\label{eq:variation}
\eeq
The corresponding action takes the form
\beq
 S ~=~ S_{(2,0,\cdots,0)} + S_{(0,2,\cdots,2)} ~=~ \frac{1}{2} \int dt d^d x \, \Big[ |\p_t \phi^a|^2 + | g^{ab} \omega_{b \, c_1 \cdots c_d} \p_1 \phi^{c_1} \cdots \p_d \phi^{c_d} |^2 \Big],
 \label{eq:action}
\eeq
where $|\cdot|$ denotes the norm with respect to the metric $g_{ab}$. 
As we have discussed in the previous section, 
this type of action can have marginally stable instanton solutions. 
Their topological charge is given by
\beq
T ~=~ \int dt \, d^d x \, \omega_{a \, b_1 \cdots b_{d}} \p_t \phi^{a} \p_{x_1} \phi^{b_1}  \cdots \p_{x_d} \phi^{b_d} 
~=~ \int_{\R^{d+1}} \phi^\ast \omega, 
\eeq
where $\phi^\ast \omega$ denotes the pullback of $\omega$ 
with respect to the map $\phi$ from the $(d+1)$-dimensional spacetime to the target space $\mathcal M$.  

One of the most significant properties of the action Eq.\,\eqref{eq:action} is that 
it is invariant under the $z=d$ Lifshitz scaling transformation
\beq
t \rightarrow \lambda^d t, \hs{10} x_i \rightarrow \lambda x_i, 
\eeq
and the spatial volume-preserving diffeomorphism
\beq
x_i \rightarrow x_i'(x), \hs{10} \det \left( \frac{\p x_i'}{\p x_j} \right) = 1. 
\eeq
As a consequence of these symmetry, 
instantons in this class of theories have continuous degeneracy
associated with the symmetry broken by the configurations. 
The corresponding moduli parameters determines the shape of the instanton
which can vary under the above transformations. 

The action of the form Eq.\,\eqref{eq:action} has some exotic properties: 
for example, it vanishes for an arbitrary static configuration 
which does not depend on one of the spatial coordinates. 
To obtain a physically reasonable model, 
we should modify the action by adding terms 
which would break the stability of the instantons. 
In the presence of such a modification, instantons are no longer true minima of the action. 
Nevertheless, the path integral in the modified model is dominated 
by the so-called constrained instantons \cite{Affleck:1980mp}
whose leading order forms are given by the BPS configurations in the original model. 
Therefore, it is still important to know the properties of the instantons in the model Eq.\,\eqref{eq:action}
and we will mainly focus on them in the following.

%%%%%%%%%%%%%%%%%%%%%%%%%%%%%%%%%%%%%%%%%%%%%%
\subsection{Anisotropic BPS baby Skyrmions}
\label{sec:baby}
Let us first see the simplest example, the sigma model lump, 
in the Lorentz symmetric K\"ahler sigma model in two dimensions 
\cite{Polyakov:1975yp}. 
Then, we will see that it takes an anisotropic shape with a fixed size
in the presence of additional potential and spatial higher derivative terms. 

When the target space $\mathcal M$ is a K\"ahler manifold, 
the standard sigma model action 
\beq
S = \int dt dx \, g_{a \bar{b}} (\p_t \phi^a \p_t \bar{\phi}^b + \p_x \phi^a \p_x \bar{\phi}^b ), 
\eeq
can be obtained by setting
\beq
W = \int dx \, \frac{i}{2} \left( \p_x \phi^a \frac{\p}{\p \phi^a} - \p_x \bar{\phi}^a \frac{\p}{\p \bar{\phi}^a} \right) K, 
\eeq
where $K$ is the K\"ahler potential which gives the target space metric $g_{a \bar{b}} = \frac{\p^2 K}{\p \phi^a \p \bar{\phi}^b}$. 
For this superpotential, the BPS equation takes the form 
\beq
\p_t \phi^a + g^{a \bar{b}} \frac{\delta W}{\delta \bar{\phi^b}} ~=~ (\p_t + i \p_x ) \phi^a ~=~ 0.
\label{eq:BPSlump}
\eeq
This is known as the BPS equation which describes the sigma model lumps. 
The corresponding topological charge is given by the pullback of the K\"ahler form \cite{Polyakov:1975yp} 
\beq
T = \int_{\R^2} \phi^\ast \left(i g_{a \bar{b}} d \phi^a \wedge d \bar{\phi}^b \right).
\eeq
Since the solution to the BPS equations is an arbitrary holomorphic map $\phi^a = \phi^a(z),~(z=t+ix)$
we can freely change the scale $\phi^a(z) \rightarrow \phi^a(\lambda z)$
without changing the value of the action. 
This continuous parameter $\lambda$ is the size modulus
associated with the scale invariance of the sigma model action.

\begin{figure}[h]
\centering
\includegraphics[width=80mm]{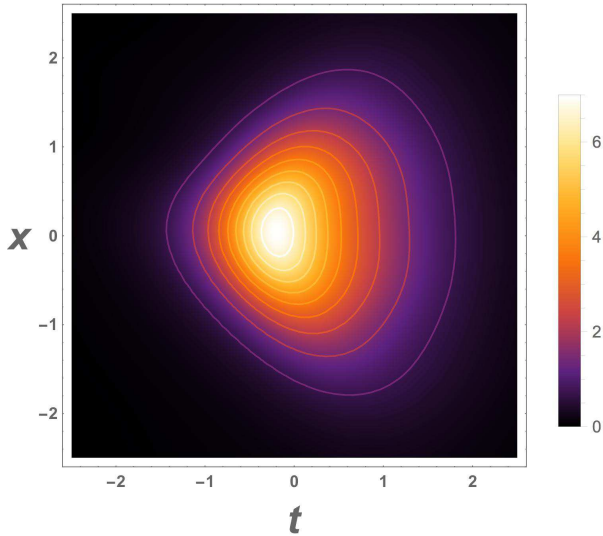}
\caption{The anisotropic BPS baby Skyrmion (numerical solution): the action density for $\alpha = \frac{1}{4}$, $\beta=\frac{1}{2}$.}
\label{fig:babyskyrmion}
\end{figure}

As we have discussed in the previous section, 
the size of the BPS lump is fixed if we turn on a potential term and higher derivative terms. 
The baby Skyrmion is such a soliton with a fixed size in the Lorentz symmetric $\C P^1$ sigma model. 
In our setup, we can introduce spatial higher derivative terms $S_{(0,n)}~(n>2)$ without breaking the BPS properties.
An example of such model can be obtained by deforming $W$ as
\beq
W = \int dx \, \bigg[ \frac{i}{2} \left( \p_x \phi^a \frac{\p}{\p \phi^a} - \p_x \bar{\phi}^a \frac{\p}{\p \bar{\phi}^a} \right) K + \alpha \, g_{a \bar{b}} \p_x \phi^a \p_x \bar{\phi}^b + w (\phi,\bar{\phi}) \bigg],
\eeq
where $\alpha$ is a constant parameter and $w(\phi, \bar{\phi})$ is a function on $\mathcal M$.
Then, the BPS equation becomes
\beq
\left( \p_t + i \p_x - \alpha \D_x \p_x \right) \phi^a = - g^{a \bar{b}} \frac{\p w}{\p \bar{\phi}^b},
\eeq
where $\D_x \p_x \phi^a = \p_x^2 \phi^a + \Gamma^a_{bc} \p_x \phi^b \p_x \phi^c$. 
Note that the scaling transformation $(t,x) \rightarrow (\lambda t, \lambda x)$ is 
no longer the symmetry of the BPS equation. 
Fig.\,\ref{fig:babyskyrmion} shows a numerical solution of the anisotropic BPS configuration 
in the $\C P^1$ model with the K\"ahler potential $K = \log (1+|\phi|^2)$ and the superpotential 
\beq
W = \int dx \left[ \frac{i}{2} \frac{\bar{\phi} \p_x \phi - \phi \p_x \bar{\phi}}{1+|\phi|^2} + \frac{ \alpha |\p_x \phi|^2 + \beta}{(1+|\phi|^2)^2} \right].
\eeq
As shown in Fig.\,\ref{fig:babyskyrmion}, the Lifshitz-type instanton has an anisotropic action density. 

%%%%%%%%%%%%%%%%%%%%%%%%%%%%%%%%%%%%%%%%%%%%%%
\subsection{Anisotropic BPS Skyrmions}
\label{sec:Skyrme}
Next, we discuss the three dimensional Lifshitz-type non-linear sigma model 
whose target space is a Lie group $\mathcal M = G$. 
Since $\pi_3(G) \cong \Z$ for any compact simple Lie group, 
there can be instantons characterized by the homotopy group $\pi_3(G)$. 
Such instantons are BPS configurations in the model with the superpotential 
which takes the form of the Wess-Zumino-Witten term 
\cite{Witten:1983ar} 
\beq
W = \frac{1}{3} \int d^2 x  \int_{-\infty}^t dt \, i \epsilon_{IJK} 
\tr \Big[ \left( i U^\dagger \p_I U \right) \left( i U^\dagger \p_J U \right) \left( i U^\dagger \p_K U \right) \Big], \hs{5} (I,J,K=0,1,2).
\eeq
This superpotential gives the following action with four spatial derivatives
\beq
S ~=~ \frac{\xi}{2} \int dt \, d^2 x \, \tr \bigg\{ (i U^\dagger \p_t U)^2 - \big[ i U^\dagger \p_1 U , i U^\dagger \p_2 U \big]^2 \bigg\},
\eeq
where $\xi$ is the inverse coupling constant. 
Since $\xi$ is dimensionless, this model is expected to be renormalizable according to the power-counting argument. 
The BPS equation takes the form
\beq
i U^\dagger \p_t U = i [ i U^\dagger \p_1 U , i U^\dagger \p_2 U ].
\label{eq:SkyrmBPS}
\eeq
This system is invariant under the $z=2$ Lifshitz scaling and 
the spatial volume-preserving diffeomorphism. 
The topological charge is given by  
\beq
T = \frac{\xi}{3} \int dt \, d^2 x \, i \epsilon^{IJK} \tr \Big[ (i U^\dagger \p_I U)(i U^\dagger \p_J U)(i U^\dagger \p_K U) \Big].
\eeq
We call the instanton in this model ``anisotropic Skyrmion" 
since it is characterized by the same topological charge (baryon number) as the Skyrmion 
in the Lorentz symmetric theories.

%%%%%%%%%%%%%%%%%%%%%%%%%%%%%%%%%%%%%%%%%%%%%%
\subsubsection{$G=SU(2)$}
\begin{figure}[h]
\begin{center}
\includegraphics[width=160mm]{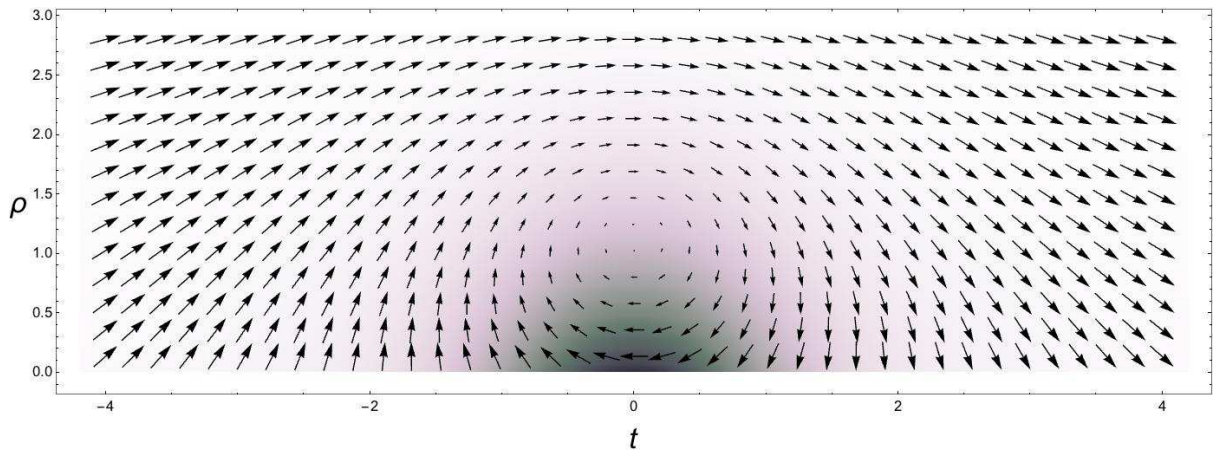}
\caption{A numerical solution. The grayscale shows the action density and 
each vector indicates the scalar field $({\rm Re} \, \phi, {\rm Im} \, \phi)$ at each point.
The action density is localized at $t=0$ on the $t$-axis.}
\label{fig:skyrmion}
\end{center}
\end{figure}

As an example, let us consider the case of $G=SU(2)$.
Let $(r,\theta)$ be polar coordinates of two-dimensional space.
Combining the spatial rotation $SO(2)$ with the $U(1)_V \subset G_L \times G_R$ symmetry, 
we can construct the following $U(1)$ symmetric ansatz 
\beq
U =
V^\dagger 
\ba{cc}
\phi & - \sqrt{1-|\phi|^2} \\
\sqrt{1-|\phi|^2} & \bar{\phi} 
\ea
V, \hs{13}
V = \exp \left( i k \theta \, \frac{\sigma_3}{2} \right),
\label{eq:Skyrmansatz}
\eeq
where $k$ is an integer and $\phi$ is a complex-valued function which is independent of the spatial angle coordinate $\theta$.\footnote{
The same ansatz has been considered in 
Skyrmions trapped inside a vortex
\cite{Gudnason:2014hsa,Kobayashi:2013aza} 
and fractional instantons on ${\mathbb R}^2 \times S^1$
with a twisted boundary condition 
\cite{Nitta:2015tua,Nitta:2014vpa}.
}
Plugging this ansatz into the BPS equation Eq.\,\eqref{eq:SkyrmBPS}, 
we obtain the following equation for $\phi$: 
\beq
\left[ \p_t + i \p_\rho + \frac{i}{2}( \phi \, \p_\rho \bar{\phi} - \bar{\phi} \, \p_\rho \phi) \right] \phi = 0,
\label{eq:Skyrmion_reduced}
\eeq
where we have defined the spatial radial coordinate by $\rho = r^2/4k = (x_1^2+x_2^2)/4k$. 
Note that in this notation, the $z=2$ Lifshitz scaling transformation becomes
\beq
t \rightarrow \lambda t, \hs{10} \rho \rightarrow \lambda \rho.
\eeq
The scalar field $\phi$ can be viewed as a map 
from the $(t,\rho)$-plane (half plane $\rho \geq 0$) to the unit disk $|\phi| \leq 1$. 
If we assume that $\phi$ takes a fixed value at infinity,
the $(t,\rho)$-plane can also be regarded as 
a disk whose boundary corresponds to the line $\rho=0$. 
Since the angle coordinate $\theta$ in the ansatz Eq.\,\eqref{eq:Skyrmansatz} 
is ill-defined on the axis of the cylindrical coordinate system $(t,\rho,\theta)$, 
the scalar field $\phi$ should satisfy the boundary condition 
\beq
\lim_{\rho \rightarrow 0} |\phi|^2 = 1. 
\eeq
Then, the topological charge reduces to
\beq
T ~=~ 2 \pi k i \xi \int_{\rho \geq 0} d ( \phi d \bar{\phi} - \bar{\phi} d \phi ) ~=~ 4 \pi k \xi \int_{\rho = 0} d \arg \phi.
\label{eq:topred}
\eeq
This topological charge counts the winding number of the map $\phi$ 
from the line $\rho=0$ to the boundary $|\phi|=1$. 
Fig.\,\ref{fig:skyrmion} shows a numerical solution of the BPS equation \eqref{eq:Skyrmion_reduced}. 
There is a zero of $\phi$ on the $(t,\rho)$-plane around which $\arg \phi$ has a winding number. 
Note that the winding number and the integer $k$ are not individually identified with the topological invariant. 

Once a BPS solution is obtained, 
we can find a continuous family of solutions by using the symmetries. 
The time translation and the Lifshitz scaling shift the zero of $\phi$, 
and hence its position on the $(t,\rho)$-plane can identified 
with a complex moduli parameter of the solution. 
We can also use the volume-preserving diffeomophism to generate a set of solutions. 
Let us focus on the $SL(2,\R)$ subgroup defined by
\beq
\ba{c} x_1 \\ x_2 \ea \rightarrow \ba{cc} a & b \\ c & d \ea \ba{c} x_1 \\ x_2 \ea, \hs{10}  ad - bc =1. 
\label{eq:SL2}
\eeq 
Since the action of the spatial rotation $SO(2) \subset SL(2,\R)$ is trivial for a rotationally symmetric solution, 
the $SL(2,\R)$ orbit of the solution is the hyperbolic plane $H^2 \cong SL(2,\R)/SO(2)$. 
Therefore, the set of solutions has two real parameters corresponding to the coordinates of the hyperbolic plane.

\paragraph{Deformation \\}
We can also add the following term without breaking the $z=2$ Lifshitz scaling invariance   
\beq
\Delta W = \frac{\alpha}{2} \int d^2x \, \tr \left( i U^\dagger \p_i U \right)^2.
\eeq
The BPS equation becomes 
\beq
i U^\dagger \p_t U = ( \epsilon_{ij} + \alpha \delta_{ij} ) \p_i ( i U^\dagger \p_j U).
\eeq
For $\alpha \not = 0$, 
there is no longer the spatial volume-preserving diffeomorphism, 
and hence the $H^2$ moduli parameters are fixed. 
On the other hand, the zero of $\phi$ remains a complex moduli parameter
since the time translation and the $z=2$ Lifshitz scaling symmetry are still unbroken. 

We remark that the superpotential $W$ now takes the form of the Wess-Zumino-Witten action. 
As pointed out in \cite{Horava:2008ih}, 
the ground state wavefuncitonal in the $(2+1)$ dimensional theory
reproduces the partition function of the 2d theory defined by the action $W$. 
A detailed study of this model would also be interesting. 

\paragraph{Compactification \\}
Let us next discuss instantons at finite temperature by compactifying the Euclidean time direction. 
In this case, we can consider an ansatz which is invariant under a combination of the time translation and 
the $U(1)$ internal symmetry $\phi \rightarrow e^{i \alpha} \phi$. 
The exact BPS solution which has such a symmetry is given by
\beq
\phi = \exp \left( 2 \pi i n \frac{t + i \rho}{\beta} \right), \hs{10} n \in \Z.
\eeq
The topological charge of this solution is independent of the inverse temperature $\beta$
\beq
T ~=~ \frac{8\pi^2 n^2 \xi}{\beta^2}\int_{0}^\beta dt \int d^2 x \, \exp \left[ - \frac{n \pi}{k \beta} (x_1^2 + x_2^2) \right] 
~=~ 8 \pi^2 n k \xi. 
\label{eq:topdensity}
\eeq

\begin{figure}[htbp]
\begin{center}
\begin{tabular}{c}
\begin{minipage}{0.30\hsize}
\begin{center}
\includegraphics[width=5cm]{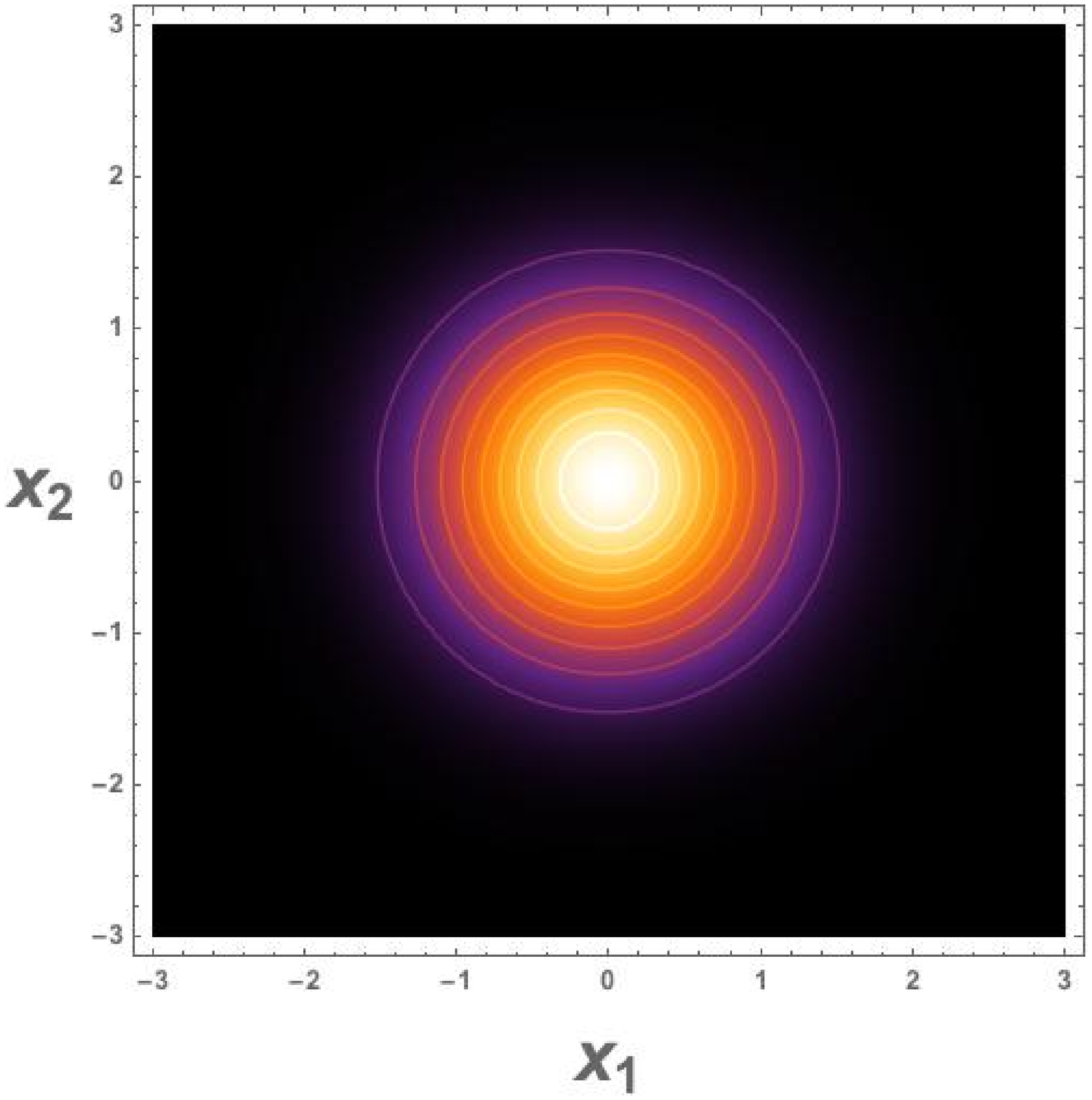}
\hs{15} 
(a) $\vartheta=0$
\end{center}
\end{minipage}
\begin{minipage}{0.30\hsize}
\begin{center}
\includegraphics[width=5cm]{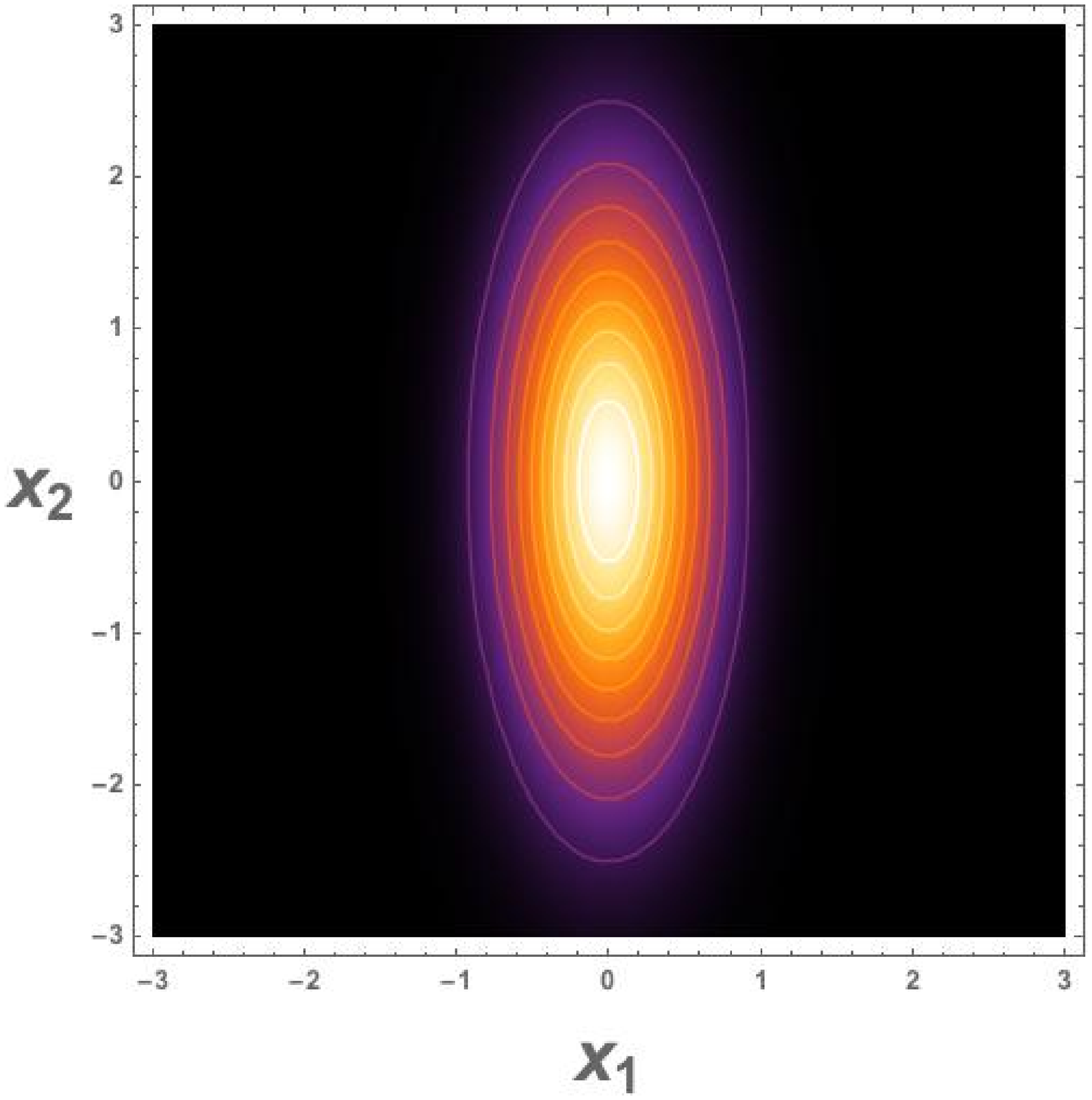}
\hs{15} 
(b) $\vartheta=1,\,\varphi=0$
\end{center}
\end{minipage}
\begin{minipage}{0.30\hsize}
\begin{center}
\includegraphics[width=5cm]{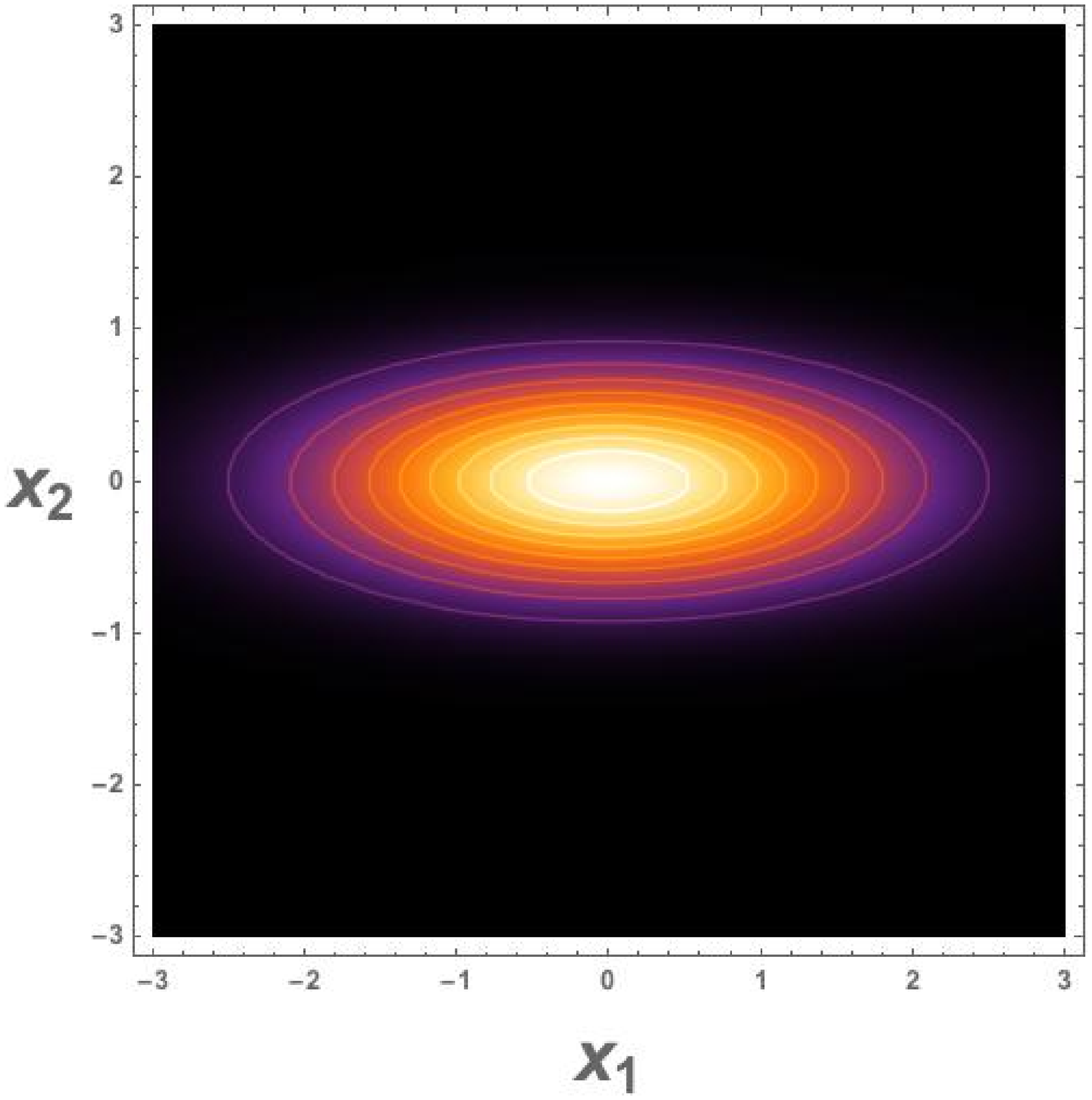}
\hs{15} 
(c) $\vartheta=1,\,\varphi=\pi$
\end{center}
\end{minipage}
\begin{minipage}{0.1\hsize}
\begin{center}
\includegraphics[width=1cm]{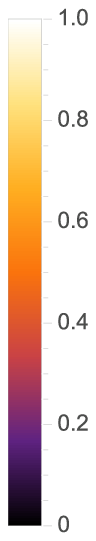}
\end{center}
\vs{5}
\end{minipage}
\end{tabular}
\caption{The action density on the $(x_1,x_2)$-plane}
\label{fig:Skyrmion_shape}
\end{center}
\end{figure}

Since we have compactified the time direction, 
the Lifshitz scaling $t \rightarrow \lambda^2 t,~x_i \rightarrow \lambda x_i$ is no longer a symmetry 
but a transformation which relates the solutions for different values of the inverse temperature $\beta$. 
On the other hand, there is still the the $SL(2,\R)$ symmetry Eq.\,\eqref{eq:SL2}, 
under which the topological charge density in Eq.\,\eqref{eq:topdensity} 
transforms in such a way that $x_1^2 + x_2^2$ is replaced by
\beq
x_1^2 + x_2^2 &\rightarrow& (x_1^2 + x_2^2) \cosh \vartheta + 
(2 x_1 x_2) \sinh \vartheta \sin \varphi + (x_1^2 - x_2^2) \sinh \vartheta \cos \varphi.
\eeq
where $(\vartheta, \varphi)$ are moduli parameters corresponding to $SL(2,\R)/SO(2)$. 
Fig.\,\ref{fig:Skyrmion_shape} shows the shape of the instanton for various values of the moduli parameters. 
The contour lines of the action density are ellipses and 
the parameters $(\vartheta,\,\varphi)$ are related to the eccentricity $e=\sqrt{1-e^{-2\vartheta}}$ 
and the rotational angle, respectively.

The exact solution discussed above is symmetric under the time translation. 
In general, such a symmetric configuration takes the form
\beq
U(t,x) ~=~ e^{ i v_L t } U(x) e^{- i v_R t }, 
\eeq
where $v_L$ and $v_R$ are fixed elements of the Lie algebras $\mathfrak{g}_L$ and $\mathfrak{g}_R$, respectively.
For this ansatz, Eq.\,\eqref{eq:SkyrmBPS} becomes
\beq
v_R - U^\dagger v_L U = i \left[ i U^\dagger \p_1 U, i U^\dagger \p_2 U \right].
\eeq
This is the BPS equation in the two dimensional system described by the action 
\beq
S = \frac{\xi}{2} \int d^2 x \, \tr \left\{ (v_R - U^\dagger v_L U )^2 - \left[ i U^\dagger \p_1 U, i U^\dagger \p_2 U \right]^2 \right\}.
\eeq
This action is bounded below by the topological charge
\beq
T ~=~ \xi \int_{\R^2} d \, \Big\{ \tr ( i v_L U d U^\dagger) - \tr( i v_R U^\dagger d U) \Big\}.
\label{eq:global_vortex}
\eeq
If the vacua $\mathcal M_{vac} = \{ U \in G ~|~ v_R = U^\dagger  v_L U \}$
has a non-trivial first homotopy group $\pi_1( \mathcal M_{vac} ) \not = \mathbf 1$, 
there can be global vortex configurations characterized by the boundary condition \cite{Nitta:2015tua}
\beq
\lim_{r \rightarrow \infty} U(x) = U_0 \, e^{i v \theta}, 
\eeq
where $U_0 \in G$ is a solution of the vacuum condition $v_R = U^\dagger  v_L U$ and 
$v$ is an element of $\mathfrak{g}_R$ such that $[v,v_R]=0$ and $e^{2\pi v i} = \mathbf 1$.
For a BPS configuratioin satisfying the above boundary condition, 
the topological charge Eq.\,\eqref{eq:global_vortex} is given by
\beq
T = 4 \pi \xi \, \tr(v_R v). 
\eeq
Note that although the action (energy) for a global vortex configuration 
is logarithmically divergent in the presence of the standard kinetic terms 
\cite{Nitta:2015tua},
it is finite in this type of model which has only potential and higher derivative terms. 

%%%%%%%%%%%%%%%%%%%%%%%%%%%%%%%%%%%%%%%%%%%%%%
\subsection{Instantons in higher dimensional sigma models}
\label{sec:higher}
\subsubsection{$O(d+1)$ sigma model}
In the previous section, we have considered 
the non-linear sigma model with $\mathcal M = SU(2) \cong S^3$ in three dimensions. 
As a generalization to higher dimensions,  
we can consider the $(d+1)$-dimensional models with $\mathcal M = S^{d+1}$, 
i.e. the $O(d+1)$ sigma model.   
The action which admits BPS instantons takes the form
\beq
S = \frac{\xi}{2} \int dt \, d^d x  \, \left[ \p_t \vec n \cdot \p_t \vec n + \vec \Omega \cdot \vec \Omega \right], \hs{10}
\Omega^a = \epsilon^{a \, b \, c_1 \cdots c_d} \, n^b \p_1 n^{c_1} \cdots \p_d n^{c_d},
\eeq
where $\vec n$ is a unit vector $\vec n \cdot \vec n = 1$.  
The Bogomol'nyi bound for this action is given 
by the following topological charge corresponding to $\pi_{d+1}(S^{d+1}) = \Z$
\beq
S ~\geq~ T ~=~ \xi \int dt d^d x \, \p_t \vec n \cdot \vec \Omega ~=~ \xi \int_{\R^{d+1}} \phi^\ast \, {\rm Vol}(S^{d+1}),
\eeq
where $\phi^\ast \, {\rm Vol}(S^{d+1})$ is the pullback of the volume form on $S^{d+1}$. 
The BPS equation is
\beq
\p_t \vec n ~=~ \vec \Omega.
\label{eq:BPS_O(d+1)}
\eeq
Let us assume that the single instanton solution is symmetric under the spatial rotation $SO(d)$. 
Combining with the $SO(d)$ symmetry of the target space, 
we can find the following hedgehog ansatz 
\beq
n^a = \sqrt{1-|\phi|^2} \, \frac{x^a}{r} ~~(a=1,\cdots,d), \hs{10} n^{d+1}+ i n^{d+2} = \phi, 
\eeq
where $r$ is the spatial radial coordinate.  
The condition for the smoothness of the configuration at $r=0$ implies that 
the complex-valued function $\phi$ should satisfy the boundary condition
\beq
\lim_{r \rightarrow 0} |\phi| = 1. 
\eeq
Substituting the ansatz into Eq.\,\eqref{eq:BPS_O(d+1)}, 
we obtain the BPS equation for $\phi$
\beq
\p_t \phi = - (1-|\phi|^2)^{\frac{d-2}{2}} \bigg[ i \p_\rho + \frac{i}{2} ( \phi \, \p_\rho \bar{\phi} - \bar{\phi} \, \p_\rho \phi ) \bigg] \phi, 
\eeq
where we have redefined the spatial radial coordinate by $\rho = r^d/d$. 
Note that this BPS equation reduces to Eq.\,\eqref{eq:Skyrmion_reduced} for $d=2$. 
The topological charge becomes 
\beq
T ~=~ \frac{V_{d-1}}{d} \xi \int_{\rho = 0} d \arg \phi . 
\eeq
As in the case of $\mathcal M = SU(2)$, this topological charge counts the winding number 
of the map from the line $\rho = 0$ to the boundary of the disk $|\phi|=1$. 

\paragraph{Compactification \\}
Assuming that the Euclidean time direction is compact, 
we can obtain the following implicit form of an exact solution 
which is symmetric under the time translation
\beq
\phi = \sqrt{1-f(\rho)^2} e^{\frac{2\pi i t}{\beta}}, \hs{10} 
\sqrt{(1- f) (1 + f)^{(-1)^d}} = \exp \left(- \frac{2\pi}{\beta} \rho - \sum_{l=1}^{\lfloor \frac{d-1}{2} \rfloor} \frac{f^{d-2l}}{d-2l} \right),
\eeq
where $\lfloor x \rfloor$ denotes the largest integer not greater than $x$. 
For any $d$, the scalar field $\phi$ has the asymptotic form
\beq
\phi \simeq  \exp \left[ \frac{2\pi i}{\beta} ( t + i \rho ) \right], \hs{10} (\rho \rightarrow \infty). 
\eeq
The scalar field $\phi$ vanishes at the spatial infinity, 
so that $\vec{n}|_{\rho \rightarrow \infty}$ defines a map 
from $S^{d-1}_{\infty}$ to $S^{d-1} = \{ \, \vec n \in S^{d+1} ~|~ \phi = 0 \, \}$.
Therefore, this configuration is a global type soliton characterized by $\pi_{d-1}(S^{d-1})$. 

%%%%%%%%%%%%%%%%%%%%%%%%%%%%%%%%%%%%%
\subsubsection{Other examples}
Another class of higher dimensional models 
which admits BPS instantons is the K\"ahler sigma models. 
Since any K\"ahler manifold has the closed forms 
\beq
\omega^{l} = ( i g_{a \bar{b}} d\phi^a \wedge d \bar{\phi}^b )^{l}, ~~~(l=1,\cdots, {\rm dim}_{\C} \mathcal M),
\eeq 
there can be instantons characterized by the topological charge of the form
\beq
T = \xi \int_{\R^{d+1}} \phi^\ast \omega^{n+1}, \hs{10} (d = 2n+1).
\eeq
This topological charge gives the BPS bound for the action 
\beq
S = \frac{\xi}{2} \int dt d^d x \, g_{a \bar{b}} \left[ \p_t \phi^a \p_t \bar{\phi}^b + (V^i \p_i \phi^a) \overline{(V^i \p_i \phi^b)} \right],
\eeq
where $V^i$ is defined by
\beq
V^i = \epsilon_{i \, j_1 \cdots j_{2n}} (\phi^\ast \omega)_{j_1 j_2} \cdots (\phi^\ast \omega)_{j_{2n-1} \, j_{2n}}, \hs{10}
(\phi^\ast \omega)_{ij} = i g_{b \bar{c}} \p_{i} \phi^{a} \p_{j} \bar{\phi}^{b}. 
\eeq
The BPS equation 
\beq
(\p_t + i V^i \p_i) \phi^a = 0, 
\eeq
can be regarded as a generalization of that in (1+1)-dimensional K\"ahler sigma models in Eq.\,\eqref{eq:BPSlump}.

When the target space $\mathcal M$ is a Calabi-Yau manifold of complex dimension $d+1$,
(the real part of) the holomorphic $(d+1)$-form can also be used as the superpotential. 
In general, any calibration form on a calibrated manifold can be used to construct a non-linear sigma model with BPS instantons. 
It would also be interesting to consider instantons corresponding 
to closed forms generated by the triplet of K\"ahler forms on hyper-K\"ahler manifolds.  

%%%%%%%%%%%%%%%%%%%%%%%%%%%%%%%%%%%%%%%%%%%%%%
\section{Instantons in Lifshitz-type gauge theories}
\label{eq:gauge}
In this section, we discuss instantons in the Lifshitz-type gauge theories. 
As is well-known, there exist instanton solutions in the pure Yang-Mills theory in the $(3+1)$-dimensional spacetime. 
In the following, we generalize the Yang-Mills instantons 
to the $(d+1)$-dimensional gauge theories with the classical $z=d-2$ Lifshitz scaling symmetry. 

%%%%%%%%%%%%%%%%%%%%%%%%%%%%%%%%%%%%%%%%%%%%%%
\subsection{Derrick's scaling argument in Lifshitz-type gauge theories}
\label{sec:Derrick_gauge}
First, we discuss the stability of instantons by generalizing 
the scaling argument in Sec.\,\ref{sec:Derrick} to the case of gauge fields. 
Since gauge fields are associated with covariant derivatives, 
we assign the scaling weights so that each component $A_i$ has the same scaling properties as the derivative $\p_i$. 
Then, the covariant derivatives transform under the scaling Eq.\,\eqref{eq:Gscaling} as
\beq
\p_i + i A_i \rightarrow \lambda^{-w_i} (\p_i + i A_i).
\eeq
If an action for the gauge field have non-trivial extrema, 
it should be stable under the scaling transformation with arbitrary weights. 
In four dimensions, the electric and the magnetic parts of the Yang-Mills action $S_{YM}$
\beq
S_{E_i} = \frac{1}{2g^2} \int dt d^3 x \, E_i^2, \hs{10} S_{B_i} = \frac{1}{2g^2} \int dt d^3 x \, B_i^2,
\label{eq:em}
\eeq
have the scaling weights with the opposite signs 
$w(S_{E_i}) = - w(S_{B_i}) = w_0 + w_i - \sum_{j=1,j \not = i}^3 w_j$ for generic $w_i~(i=0,1,2,3)$,
whereas both terms are scale invariant for $w_0=w_i~(i=1,2,3)$. 
Therefore, the Yang-Mills instanton is stable under the generic scaling    
and marginally stable for the scaling with $w_0=w_i$. 

As a generalization to higher dimensions, 
we can consider an analogue of Eq.\,\eqref{eq:em}. 
For $d=2n+1$, define $E_i$ and $B_i$ by
\beq
E_i = F_{t i}, \hs{10} B_i = \frac{1}{2^n} \epsilon_{i \, j_1 \cdots j_{2n}} F_{j_1 j_2} \cdots F_{j_{2n-1} \, j_{2n}}. 
\eeq
Since $S_{E_i}$ and $S_{B_i}$ have the opposite weights
\beq
w(S_{E_i}) = - w(S_{B_i}) = w_0 + w_i - \sum_{j=1, \, j \not = i}^d w_j,
\eeq
there can be instanton solutions in the higher dimensional gauge theory described by
\beq
S = \frac{1}{2g^2} \int dt d^d x \, \tr \left( E_i^2 + B_i^2 \right).
\label{eq:gaugeaction}
\eeq
Since this action is invariant under the $z=d-2$ Lifshitz scaling ($w_0 = d-2,~w_i=1~(i=1,\cdots,d)$), 
the instantons are marginally stable and hence they have associated size moduli parameters. 
Note that the gauge coupling constant $g$ is dimensionless and 
hence this model is expected to be renormalizable according to the power-counting argument. 

%%%%%%%%%%%%%%%%%%%%%%%%%%%%%%%%%%%%%%%%%%%
\subsection{Weyl rescaling and coset space dimensional reduction}
\label{sec:red}
An important property of the action Eq.\,\eqref{eq:gaugeaction} is that 
it is invariant under a generalized version of the Weyl transformation. 
Let us put the model on a curved space with a metric of the form
\beq
ds^2 = N^2 dt^2 + g_{ij} (dx^i + N^i dt) (dx^j + N^j dt).
\eeq
The covariantized action\footnote{
The action which has covariance 
under the foliation-preserving diffeomorphism $t \rightarrow t'(t)$ and $x_i \rightarrow x_i'(t, x_j)$ takes the form
\beq
S = \int dt d^dx \, N \sqrt{\det g_{ij}} \, g^{ij} (E_i E_j + B_i B_j), \notag
\eeq
where $E_i$ and $B_i$ are given by
\beq
E_i = \frac{1}{N} ( F_{t i} - N^j F_{ji} ), \hs{5} 
B_i = \frac{1}{2^n} \frac{1}{\sqrt{\det g_{ij}}} g_{ik} \epsilon^{k \, j_1 \cdots j_{2n}} F_{j_1 j_2} \cdots F_{j_{2n-1} j_{2n}}.
\eeq
}
is invariant under the following change of the metric 
\beq
N \rightarrow \lambda(t, x_i)^{d-2} N, \hs{10} g_{ij} \rightarrow \lambda(t,x_i)^2 g_{ij}. 
\eeq
Using this anisotropic Weyl transformation with $\lambda = r^{-1}$, 
we can map the flat spacetime to the direct product of the hyperbolic plane 
and the $(d-1)$-dimensional sphere $H^2 \times S^{d-1}$
\beq
ds^2 = dt^2 + dr^2 + r^2 d \Omega_{d-1}^2 \sim \frac{1}{(d-2)^2} \frac{dt^2 + d\rho^2}{\rho^2} + d \Omega_{d-1}^2,
\eeq
where $\rho = r^{d-2}/(d-2)$. 
Therefore, the theory can be reduced to a two-dimensional system defined on the hyperbolic plane $H^2$ 
by taking an ansatz which is symmetric under a combination of the $SO(d)$ rotation and the gauge transformation. 
It is known that the self-dual equation in the four dimensional pure Yang-Mills theory $(d=3)$ 
reduces to the vortex equation on the hyperbolic plane \cite{Witten:1976ck}. 
This type of the dimensional reduction is called the coset space dimensional reduction \cite{Forgacs:1979zs}. 

We have already seen the examples of the coset space dimensional reduction for scalar fields (zero forms) in the previous section. 
In the following, we will see that the $SO(d)$ symmetric instantons in the Lifshitz-type gauge theories
can also be mapped to vortices on the hyperbolic plane by the dimensional reduction.

%%%%%%%%%%%%%%%%%%%%%%%%%%%%%%%%%%%%%%%%%%%%%%
\subsection{Generalized Yang-Mills Instantons}
\label{sec:SU(4)}
As in the case of scalar field theories, the superpotential formalism (detailed balance condition)
can be used to construct actions which admit BPS configurations. 
The scale invariant action Eq.\,\eqref{eq:gaugeaction} for the $SU(N)$ gauge field $A$
can be obtained by using the Chern-Simons form as the superpotential
\beq
W ~=~ \frac{1}{n+1} \int_{\R^{d}} CS_{d}(A) , \hs{10} d \, CS_{d} = \tr \underbrace{( F\wedge \cdots \wedge F)}_{n+1}.
\eeq
The corresponding topological charge is given by
\beq
T ~=~ \frac{1}{g^2} \int dt d^d x \, \tr( E_i B_i ) ~=~ \frac{1}{(n+1) g^2} \int_{\R^{d+1}} \tr ( F\wedge \cdots \wedge F). 
\eeq
This gives the BPS bound for the action 
\beq
S ~\geq~ T + \frac{1}{2g^2} \int dt d^d x \, \tr \left( E_i - B_i \right)^2. 
\eeq
The BPS equation is $E_i = B_i$, that is,
\beq
F_{ti} = \frac{1}{2^n} \epsilon_{i \, j_1 \cdots j_{2n}} F_{j_1 j_2} \cdots F_{j_{2n-1} j_{2n}}.
\eeq
This is a generalization of the self-dual equation in the four dimensional Yang-Mills theory. 

%%%%%%%%%%%%%%%%%%%%%%%%%%%%%%%%%%%%%%%%%%%%%%
\subsubsection{$(5+1)$-dimensional $SU(4)$ gauge theory}
Let us consider the simplest example in the (5+1)-dimensional $SU(4)$ gauge theory, 
in which we can consider the $SO(5) \subset SU(4)$ invariant ansatz.
The action takes the form
\beq
S = \frac{1}{2g^2} \int dt d^5 x \, \tr \left[  (F_{t i})^2 + \left(\frac{1}{4} \epsilon^{ijklm} F_{jk} F_{lm} \right)^2 \right].
\label{eq:instaction}
\eeq
This model has the classical $z=3$ Lifshitz scale invariance
\beq
(\p_t + i A_t) \rightarrow \lambda^{-3} (\p_t + i A_t) , \hs{10}
(\p_i + i A_i) \rightarrow \lambda^{-1} (\p_i + i A_i) .
\eeq
The BPS equation and the topological charge are given by
\beq
F_{ti} = \frac{1}{4} \epsilon_{ijklm} F_{jk} F_{lm}, \hs{15} 
T = \frac{1}{3g^2} \int_{\R^6} \tr[ F \wedge F \wedge F ].
\eeq

Let us consider configurations which are invariant under a combination of the spatial $SO(5)$ rotation
and an $SO(5)$ subgroup in the $SU(4)$ gauge group. 
The adjoint representation $\mathbf{15}$ of $SU(4)$ can be decomposed into 
the vector representation $\mathbf{5}$ and the anti-symmetric representation $\mathbf{10}$ of the $SO(5)$ subgroup. 
Correspondingly, the $\mathfrak{su}(4)$ generators $t^a~(a=1,\cdots,15)$ are decomposed into the 5d gamma matrices and 
the $SO(5)$ generators
\beq
t^a ~\rightarrow~ \Gamma_i,~~\Gamma_{ij} = \frac{i}{2} [ \Gamma_i , \Gamma_j ], 
\eeq
where our convention for the gamma matrices is $\{ \Gamma_i , \Gamma_j\} = 2 \delta_{ij}$ 
and $\Gamma_i^\dagger = \Gamma_i$. 
Then, we can find the following basis of $SO(5)$ invariant one-forms
\beq
\omega_t = \Gamma_i \hat{x}^i dt, \hs{10} \omega_r = \Gamma_i \hat{x}^i dr, \hs{10} 
\omega_{\pm} = \Big[ \Gamma_{ij} \hat{x}^i \pm i \left( \delta_{ij} - \hat{x}_i \hat{x}_j \right) \Gamma_i \Big] dx^j,
\eeq
where $r$ is the spatial radial coordinate and $\hat x_i = x_i / r$. 
In terms of the $SO(5)$ invariant one-forms, 
the most general $SO(5)$ symmetric ansatz for the gauge field is given by
\beq
A = \frac{1}{2} \left( a_t \omega_t + a_r \omega_r + \frac{\phi-1}{2r} \omega_+ + \frac{\bar{\phi}-1}{2r} \omega_- \right),
\label{eq:instansatz}
\eeq
where $a_t, a_r$ and $\phi$ are functions of $t$ and $r$. 
For this gauge field, the field strength takes the form
\beq
F = \frac{1}{2} \left[ \Gamma_i \hat{x}^i f  
+ \frac{1}{2r} \Big( \D \phi  \wedge \omega_+ + \overline{\D \phi} \wedge \omega_-  \Big)
+ \frac{i}{8r^2} \left( |\phi|^2-1 \right) (\omega_+ \wedge \omega_- + \omega_- \wedge \omega_+) \right],
\eeq
where $f$ is the two-dimensional field strength and $\D \phi$ is the covariant derivative defined by
\beq
f = (\p_t a_r - \p_r a_t) \, dt \wedge dr, \hs{10} \D \phi = (\p_t + i a_t ) \phi  \, dt + (\p_r + i a_r) \phi \, dr.
\eeq 
Plugging into the BPS equation, we obtain the follwoing reduced equations 
for the two dimensional fields $a$ and $\phi$: 
\beq
9 \rho^2 f_{t \rho} + (|\phi|^2-1)^2 = 0, \hs{10} \D_t \phi - i (|\phi|^2-1) \D_\rho \phi = 0,
\label{eq:vortex}
\eeq
where we have redefined the spatial radial coordinate by $\rho= r^3/9$.
These equations can be regarded as the vortex equations in the following system defined on the hyperbolic plane
\beq
S = \frac{3V_4}{2g^2} \int dt d\rho \, G \left[ f_{t \rho} f^{t \rho} 
+ 4 (\D_t \phi) \overline{(\D^t \phi)} + 4 (|\phi|^2-1)^2 (\D_\rho \phi) \overline{(\D^\rho \phi)} + (|\phi|^2-1)^4 \right], 
\label{eq:2d_action}
\eeq
where $V_4$ is the volume of the unit 4-sphere and 
the spacetime indices are raised with the hyperbolic metric $G$ 
\beq
ds^2 = G (dt^2 + d \rho^2) = \frac{dt^2+d\rho^2}{9\rho^2}.
\eeq
This action can be obtained by substituting the rotationally symmetric ansatz 
Eq.\,\eqref{eq:instansatz} into the original action Eq.\,\eqref{eq:instaction}. 
Similarly, the 5d Chern-Simons term reduces to the superpotential for the two-dimensional system  
\beq
W = - \int d\rho \left[ a_\rho + \frac{i}{2} (|\phi|^2-2) \left(  \phi \D_\rho \bar{\phi} - \bar{\phi} \D_\rho \phi \right) \right]. 
\eeq
In terms of this superpotential, the two-dimensional action can be rewritten as
\beq
S = T + \frac{3V_4}{2g^2} \int dt d\rho \, \left[ G \left( G^{-1} f_{t \rho} - \frac{\delta W}{\delta a_\rho} \right)^2 + 
\left| 2 \D_t \phi - \frac{\delta W}{\delta \bar{\phi}} \right|^2 \right],
\eeq
where the topological charge is given by the first Chern number (up to trivial boundary terms)
\beq
T = - \frac{3 V_4}{g^2} \int_{H^2}  \, f.
\eeq
The two-dimensional system is anisotropic in the sense that 
the action Eq.\,\eqref{eq:2d_action} is not invariant 
under the whole $SL(2,\R)$ isometry of the hyperbolic plane. 
It is invariant under the subgroup of the isometry corresponding to the time translation and the Lifshitz scaling. 
These two symmetries shift the zeros of $\phi$ on the $(t,\rho)$-plane,
which correspond to the positions and the size moduli of the 6d instantons. 

\begin{figure}[htbp]
\begin{center}
\begin{tabular}{c}
\begin{minipage}{0.45\hsize}
\begin{center}
\includegraphics[width=7cm]{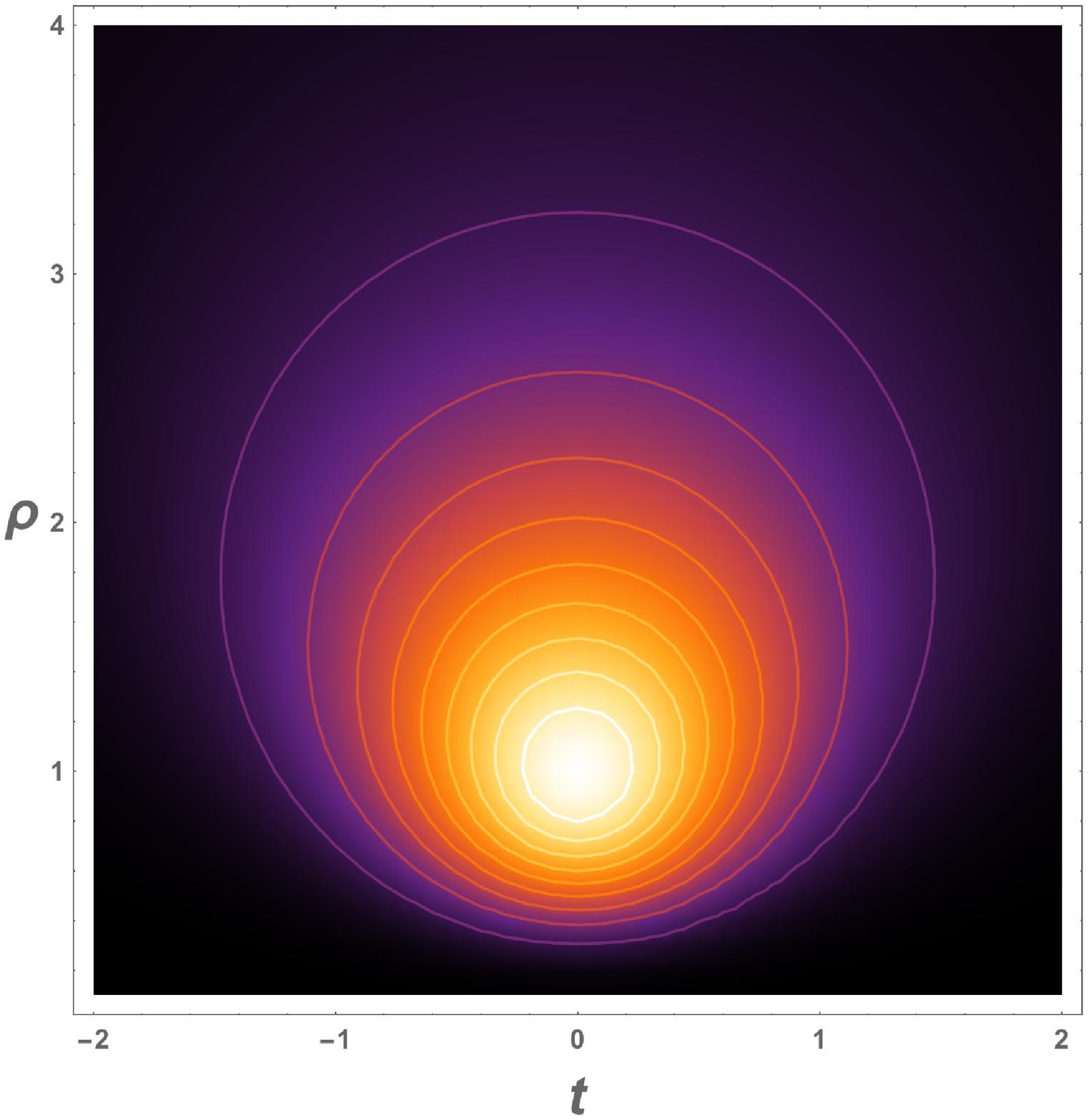}
\hs{15} 
(a) 4d instanton
\end{center}
\end{minipage}
\begin{minipage}{0.45\hsize}
\begin{center}
\includegraphics[width=7cm]{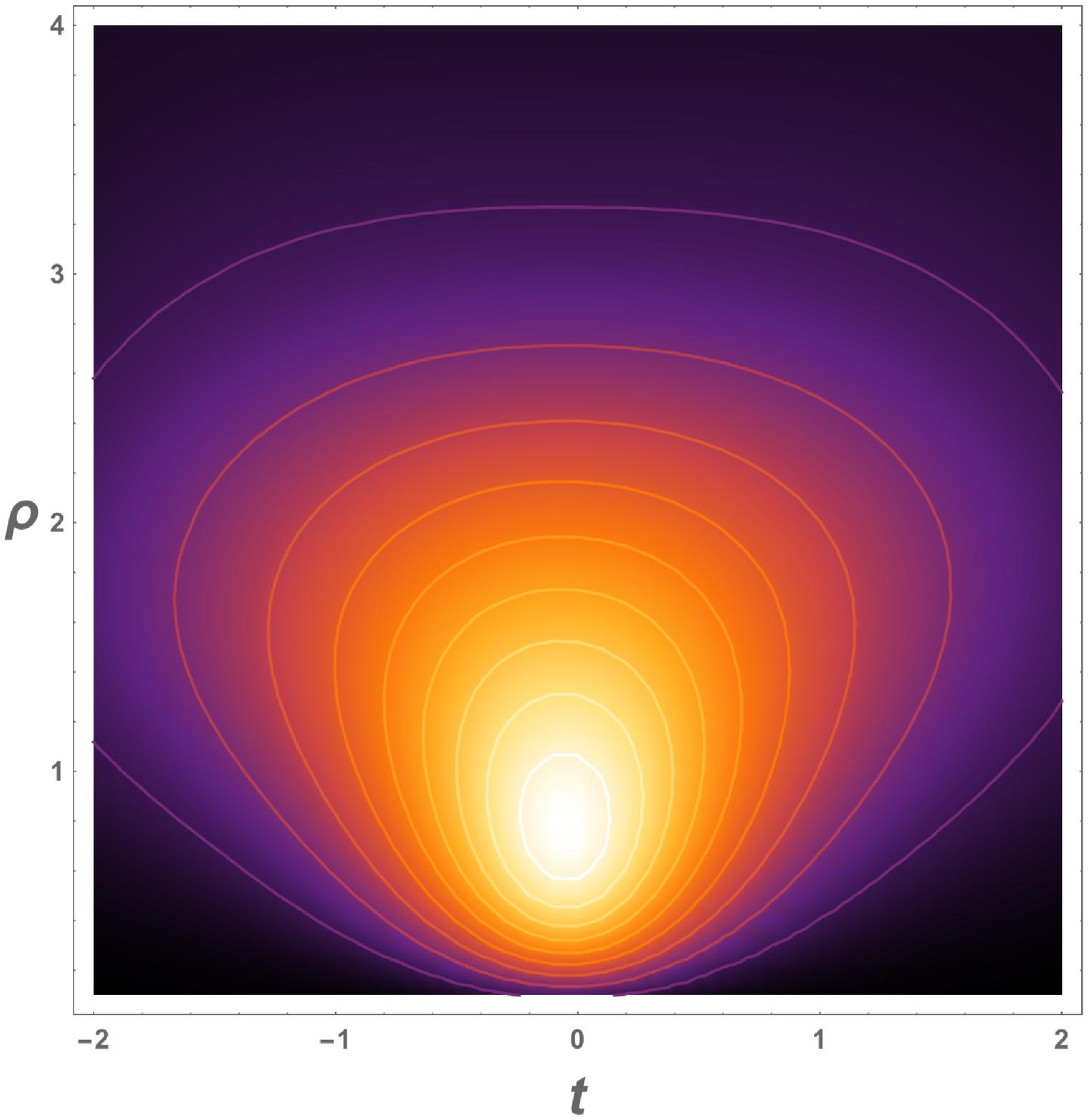}
\hs{15} 
(b) 6d instanton
\end{center}
\end{minipage}
\begin{minipage}{0.1\hsize}
\begin{center}
\includegraphics[width=1cm]{bar.eps}
\vs{5}
\end{center}
\end{minipage}
\end{tabular}
\caption{The action densities of the hyperbolic vortices (excluding the volume element of the hyperbolic plane). 
Note that the action densities for the instantons, which can be obtained by multiplying the volume element, are localized around $t=0$ on the $t$-axis.}
\label{fig:instantons}
\end{center}
\end{figure}

It has been shown in \cite{Witten:1976ck} that the $SO(3)$ invariant instantons\footnote{
The ansatz for the SO(3) invariant instantons in the 4d $SU(2)$ Yang-Mills theory is given by Eq.\,\eqref{eq:instansatz}
with $\Gamma_i$ replaced by the Pauli matrices.} 
in the 4d $SU(2)$ Yang-Mills theory 
are described by the hyperbolic vortex equations in the two-dimensional system characterized by  
\beq
G = \frac{1}{\rho^2}, \hs{10} W= - 2 \int d \rho \left[ a_\rho - \frac{i}{2} \left(  \phi \D_\rho \bar{\phi} - \bar{\phi} \D_\rho \phi \right) \right]. 
\eeq
Fig. \ref{fig:instantons} shows the action densities of the hyperbolic vortices 
corresponding to 4d and 6d instantons. 
The profile of the 4d instanton is invariant under the $SO(2)$ subgroup of the $SL(2,\R)$ isometry 
whereas that of the 6d instanton is not symmetric, reflecting the anisotropy of the original system. 

\paragraph{Compactification \\}
Next, let us consider the finite-temperature case. 
In the standard four dimensional $SU(N)$ Yang-Mills theory,  
instantons at finite temperature are called calorons \cite{Harrington:1978ve, Rossi:1978qe, Gross:1980br}.   
It is known that in the presence of a non-trivial holonomy around the Euclidean time circle, 
each caloron has $N$ constituents which can be identified with magnetic monopoles 
\cite{Lee:1997vp, Lee:1998vu, Kraan:1998kp, Kraan:1998pm, Kraan:1998sn}.
Here, we look for an analogous object which constitutes the instanton in our setup.

Since the finite-temperature system has the additional $U(1)$ symmetry corresponding to the time translation, 
we can consider an ansatz which is invariant under the $U(1) \times SO(5)$ transformations. 
By appropriately choosing the gauge, we can find the following ansatz 
\beq
\phi = \bar{\phi} = \varphi(\rho), \hs{10} a_t = a(\rho), \hs{10} a_\rho = 0. 
\eeq
Then, the BPS equations Eq.\,\eqref{eq:vortex} reduces to 
\beq
9 \rho^2 a'(\rho) = (\varphi^2-1)^2, \hs{10} \varphi'(\rho) = \frac{a \varphi}{\varphi^2-1}. 
\eeq
The smoothness of the ansatz Eq.\,\eqref{eq:instansatz} at $\rho=0$ requires that 
the profile functions $\varphi(\rho)$ and $a(\rho)$ should satisfy the boundary conditions 
\beq
\varphi(\rho = 0) = 1, \hs{10} a(\rho = 0) = 0. 
\eeq
On the other hand, $a(\infty)$ is a free parameter  
which determines the value of the Polyakov loop at the spatial infinity. 
For $a(\infty) \not = 0$, the $SU(4)$ gauge symmetry is broken to $SU(2) \times SU(2) \times U(1)$.

\begin{figure}[h]
\centering
\includegraphics[width=80mm]{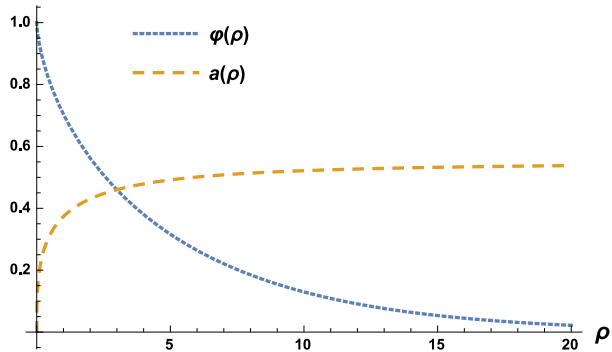}
\caption{A numerical solution}
\label{fig:monopole}
\end{figure}

The time independent solution discussed above can be identified 
with the BPS soliton in the following five dimensional system 
which can be obtained by the dimensional reduction along the periodic time direction: 
\beq
S = \frac{1}{2g^2} \int d^5 x \, \tr \Big[ (\D_i \Phi)^2 + B_i^2 \Big], \hs{10} 
B_i = \frac{1}{4} \epsilon_{ijklm} F_{jk} F_{lm},
\eeq
where $\Phi$ is the adjoint scalar field corresponding to $A_t$. 
The topological charge of the BPS soliton in this system is given by
\beq
T = \frac{1}{2g^2} \int_{\R^5} \tr( d_A \Phi \wedge F \wedge F ).
\label{eq:topGmono}
\eeq
The generalized monopole configurations characterized 
by this topological charge has been discussed in \cite{Kihara:2004yz}.

We can also perform the dimensional reduction along one of the spatial directions instead of the time direction. 
The resulting model is a five dimensional $z=3$ Lifshitz-type gauge theory with an adjoint scalar. 
It is characterized by the superpotential
\beq
W = \int_{\R^4} \tr \left( \Phi F \wedge F \right).
\eeq
The corresponding action takes the form
\beq
S = \frac{1}{2g^2} \int dt d^4 x \, \tr \left[ (F_{t i})^2 + (\D_t \Phi)^2 + \left( \frac{1}{4} \epsilon_{ijkl} F_{ij} F_{kl} \right)^2 + \left( \frac{1}{2} \epsilon_{ijkl} \{ \D_j \Phi, F_{kl} \} \right)^2 \right],
\eeq
This gauge theory is super-renormalizable since the coupling constant has the positive dimension $[g^2] = [\p_x]$.
The BPS soliton in this system has the same topological charge as Eq.\,\eqref{eq:topGmono} 
and obeys the BPS equations
\beq
\D_t \Phi = \frac{1}{4} \epsilon_{ijkl} F_{ij}F_{kl}, \hs{10} F_{ti} = \frac{1}{2} \epsilon_{ijkl} \{ \D_j \Phi, F_{kl} \}. 
\eeq
This equation can also be viewed as a generalization of the BPS monopole equation. 

%%%%%%%%%%%%%%%%%%%%%%%%%%%%%%%%%%%%%%%%%%%%%%
\section{Summary and Discussion}
In this paper, we have discussed BPS instantons in the Lifshitz-type non-linear sigma models and gauge theories. 
In the ``supersymmetric models" in which the detailed balance condition is satisfied, 
the BPS equations for the instantons are described by the gradient flow equations for the ``superpotential" $W$. 
For the Lifshitz-type non-linear sigma models, 
we have focused on BPS instantons in the $(d+1)$-dimensional settings where 
the action is invariant under the $z=d$ Lifshitz scaling and the spatial volume-preserving diffeomorphism. 
As examples, the anisotropic BPS (baby) Skyrmions and 
their higher dimensional generalizations have been discussed 
in the Lifshitz-type $O(d+1)$ sigma models. 
For the Lifshitz-type gauge theories, 
we have discussed a generalization of the Yang-Mills instanton to the higher dimensional theories. 
The anisotropic Weyl rescaling and the coset space dimensional reduction 
have been used to map 
the instantons to Abelian vortices in the anisotropic system on the hyperbolic plane.
As in the case of the instantons in four dimensions \cite{Manton:2010wu,Eto:2012aa}, 
it is also interesting to consider more general dimensional reductions 
which give non-Abelian theories in the two-dimensional spacetime 
and non-Abelian vortices in such theories.

Although the instantons in our setup are 
described by very simple BPS equations, 
the models we have considered have some exotic properties. 
For example, the action identically vanishes for any static configuration 
which is independent of one of the spatial coordinates.
It is possible to modify the action to obtain a physically reasonable model 
without breaking the BPS properties and the Lifshitz scaling invariance. 
In particular, it is interesting to consider the anisotropic Skrymions 
in the Lifshitz-type model characterized by the superpotential 
which takes the form of the Wess-Zumino-Witten model.

The (2+1)-dimensional model we have considered in Sec.\,\ref{sec:Skyrme} 
can be embedded into a supersymmetric model, 
in which the instanton configurations preserve a half of the supersymmetry. 
It should be very interesting to consider deformed supersymmetry algebras 
which allow us to compute the path integrals by means of the localization technique. 
For higher dimensions, the supersymmetry in the Lifshitz-type field theories has not yet been fully understood. 
We will discuss this topic in the near future. 

\subsection*{Acknowledgments} 
The work of T.\ F. and M.\ N.\ is supported in part by Grant-in-Aid for
Scientific Research (No. 25400268) from the Ministry of Education, Culture, Sports, Science and Technology (MEXT) of Japan.
The work of M.~N. is also supported in part by MEXT-Supported Program
for the Strategic Research Foundation at Private Universities,
``Topological Science," (grant number S1511006).

\appendix
\section{A supersymmetry in Lifshitz-type theories}
\label{appendix:SUSY}
Let us consider the follwoing SUSY algebra in $(2+1)$-dimensional spacetime
\beq
\{ Q , \bar{Q} \} = 2 i \p_t, \hs{10} Q^2=\bar{Q}^2=0.
\label{eq:algebra}
\eeq
where $Q$ and $\bar{Q}$ are supercharges 
which transform as spinors under the spacial rotation in 2d Euclidean space
\beq
Q \rightarrow e^{i \frac{\theta}{2}} Q, \hs{10} 
\bar{Q} \rightarrow e^{-i \frac{\theta}{2}} \bar{Q}.
\eeq
Note that Eq.\,\eqref{eq:algebra} is a subalgebra 
in the standard 3d $\mathcal N = 2$ SUSY algebra (4 real supercharges). 
The supercharges can be represented by the following derivatives in the superspace
\beq
Q = \frac{\p}{\p \theta} + i \bar{\theta} \p_t , \hs{10} 
\bar{Q} = \frac{\p}{\p \bar{\theta}} + i \theta \p_t.
\eeq
The supercharges anti-commute with 
\beq
D = \frac{\p}{\p \theta} - i \bar{\theta} \p_t, \hs{10}
\bar{D} = \frac{\p}{\p \bar{\theta}} - i \theta \p_t. 
\eeq
The non-linear sigma model discussed in Sec.\,\ref{sec:action} can be constructed 
in terms of the real multiplets
\beq
\Phi^a = \phi^a + \theta \psi^a + \bar{\psi} \bar{\theta}^a + \theta \bar{\theta} F^a, \hs{10} \bar{\Phi} = \Phi.
\eeq
The simplest action for the real multiplets is given by
\beq
\mathcal L = - \int d \theta d \bar{\theta} \Big[ g_{ab}(\Phi) D \Phi^a \bar{D} \Phi^b + \mathcal W(\Phi,\p_i \Phi, \cdots ) \Big],
\eeq
where $g_{ab}$ is the Riemann metric on the target space and $\mathcal W$ is a function of $\Phi$, $\p_i \Phi$ and higher spatial derivatives. 
In terms of the component fields, the kinetic part takes the standard sigma model form
\beq
\mathcal L_{\rm kin} &=& g_{ab} \Big[  \p_t \phi^a \p_t \phi^b + (F^a + \Gamma^a_{cd} \psi^c \bar{\psi}^d)(F^b + \Gamma^b_{cd} \psi^c \bar{\psi}^d) + i ( \psi^a \D_t \bar{\psi}^b - \D_t \psi^a \bar{\psi}^b) \Big] \\
&{}& - R_{abcd} \psi^a \psi^b \bar{\psi}^c \bar{\psi}^d. \notag
\eeq
The superpotential part is given by
\beq
\mathcal L_{W} = \frac{\delta W}{\delta \phi^a} F^a 
+ \mathcal W_{ab} \psi^a \bar{\psi}^b + \mathcal W_{ab}^i (\p_i \psi^a \bar{\psi}^b + \psi^b \p_i \bar{\psi}^a)
+  \mathcal W_{ab}^{ij}(\p_i \psi^a \p_j \bar{\psi}^b + \p_j \psi^b \p_i \bar{\psi}^a)
+ \cdots,
\eeq
where we have defined
\beq
&\displaystyle \frac{\delta W}{\delta \phi^a} = \frac{\p \mathcal W}{\p \phi^a} - \p_i \frac{\p \mathcal W}{\p (\p_i \phi^a )} + \cdots, & \\
&\displaystyle \mathcal W_{ab} = \frac{\p^2 \mathcal W}{\p \phi^a \p \phi^b}, \hs{10} 
\mathcal W_{ab}^i = \frac{\p^2 \mathcal W}{\p (\p_i \phi^a) \p \phi^b}, \hs{10}
\mathcal W_{ab}^{ij} = \frac{\p^2 \mathcal W}{\p (\p_i \phi^a) \p (\p_j \phi^b)}.&
\eeq
Eliminating the auxiliary fields $F^a$, we obtain the following bosonic part 
\beq
\mathcal L_{\rm bosonic} = g_{ab} \p_t \phi^a \p_t \phi^b 
- g^{ab} \frac{\delta W}{\delta \phi^a} \frac{\delta W}{\delta \phi^b}.
\eeq
This becomes the Lagrangian discussed in Sec.\,\ref{sec:action} after the Wick-rotation. 
The lower dimensional supersymmetric action can be obtained by 
the standard dimensional reduction along the spatial directions.

\end{document}